\documentstyle[preprint,aps]{revtex}
\newcommand{\bff}[1]{{\mbox{\boldmath $#1$}}}
\begin{document}
\draft
\renewcommand\baselinestretch{1.32}
\input epsf

\title{Time-odd mean fields in the rotating frame:\\
microscopic nature of nuclear magnetism.}

\author{A. V. Afanasjev\cite{LAT} and P.\ Ring}

\address{Physik-Department der Technischen Universit{\"a}t 
M{\"u}nchen, D-85747 Garching, Germany}

\date{\today}
\maketitle

\begin{abstract}
The microscopic role of nuclear magnetism in rotating frame
is investigated for the first time in the framework of the 
cranked relativistic mean field theory. It is shown that nuclear 
magnetism modifies the expectation values of single-particle spin, 
orbital and total angular momenta along the rotational axis 
effectively creating additional angular momentum. This effect 
leads to the increase of kinematic and dynamic moments of inertia 
at given rotational frequency and has an impact on effective 
alignments.
\end{abstract}
\vspace{0.5in}
\pacs{PACS numbers: 21.60.-n, 21.60.Cs, 21.60.Jz}
\vspace{0.3in}
\narrowtext

  It is a well established fact that microscopic mean field theories 
provide a good description of the properties of finite nuclei. The most
succesful among them are the non-relativistic theories based 
on the finite range forces of the Gogny type and zero-range 
forces of the Skyrme type \cite{RS.80} and relativistic mean 
field (RMF) theory \cite{R.96}. In these theories the physical observables
are most sensitive to the time-even fields, which has widely been
investigated throughout the periodic table. On the other 
hand, the properties of time-odd fields, which appear only in the 
nuclear systems with broken time-reversal symmetry, are poorly 
known. However, it is known nowadays that these fields are very 
important for the proper description of rotating nuclei 
\cite{PWM.85,KR.93,DD.95,AKR.96}, magnetic moments \cite{HR.88}, 
the properties of the $N=Z$ nuclei \cite{S.98} and pairing 
correlations \cite{RBRM.99}.

  In rotating nuclei, the time-odd mean fields emerge both
from the Coriolis operator and from the parts of the 
Hamiltonian (Lagrangian) related to the currents. Both in 
relativistic \cite{KR.93,AKR.96} and non-relativistic 
\cite{PWM.85,DD.95,YM.99} mean field theories the most 
important modification coming from the presence of currents 
lies in the renormalization of the moments of inertia. They 
increase the kinematic ($J^{(1)}$) and dynamic ($J^{(2)}$) 
moments of inertia by $\sim 20-30\%$. However, the microscopic 
mechanism of these modifications has not been well understood 
till now. 

The aim of the present article is to investigate 
this mechanism staying strictly in the framework of the
cranking model. As a theoretical tool we are using the cranked relativistic 
mean field (CRMF) theory \cite{KR.89,KR.93,AKR.96} in which 
the time-odd mean fields are defined in a unique way. 
Since we are interested in a general understanding of 
the microscopic mechanism of the renormalization of the moments 
of inertia, the pairing correlations are neglected in the 
present investigation.

 In RMF theory the nucleus is described as a system
of point-like nucleons (Dirac spinors) coupled to the mesons 
and to the photons. The nucleons interact by the exchange 
of several mesons: isoscalar scalar $\sigma$-meson, 
isoscalar vector $\omega$-meson, isovector vector $\rho$-meson 
and the photon. The transformation to the rotating frame 
in the one-dimensional cranking approximation leads 
to the CRMF theory, the details of which are given in 
Refs.\ \cite{KR.89,KR.90,KR.93,AKR.96}. The main 
features of this theory important in the present 
discussion are outlined below.

In the Hartree approximation, the stationary Dirac equation 
for the nucleons in the rotating frame is given by
\begin{eqnarray}
\left\{\bff\alpha(-i\bff\nabla-\bff V(\bff r))+
V_0(\bff r) + \beta(m+S(\bff r))- \Omega_x\hat{J}_x\right\}
\psi_i= \nonumber \\
 =  \epsilon_i\psi_i
\label{eq6}
\end{eqnarray}
where $V_{0}({\bf r})$ represents a repulsive vector potential:
\begin{equation}
V_0(\bff r)~=~g_\omega\omega_0(\bff r)+g_\rho\tau_3\rho_0(\bff r)
+e\frac{1-\tau_3}{2}A_0(\bff r)
\end{equation}
which contains time-like components of the vector mesons and 
$S({\bf r})$ an attractive scalar potential:
\begin{equation}
S(\bff r)=g_\sigma\sigma(\bff r).
\end{equation}
A magnetic potential $\bff V(\bff r)$:
\begin{equation}
\bff V(\bff r)~=~g_\omega\bff\omega(\bff r)
+g_\rho\tau_3\bff\rho(\bff r)+
e\frac{1-\tau_3}{2}\bff A(\bff r)
\end{equation}
originates from the space-like components of the vector
mesons. Note that in these equations, the four-vector 
components of the vector fields $\omega^\mu $, $\rho^\mu$ 
and $A^\mu$ are separated into the time-like ($\omega _0$, 
$\rho_0$ and $A_0$) and space-like 
[$\bff\omega\equiv(\omega^x,\omega^y,\omega^z)$,
$\bff\rho\equiv(\rho^x,\rho^y,\rho^z)$, and $\bff
A\equiv(A^x,A^y,A^z)$] components. Finally the term
\begin{equation}
{-\Omega}_x\hat{J}_x~=~
{-\Omega}_x(\hat{L}_x+\frac{1}{2}\hat{\Sigma}_x)
\end{equation}
represents the Coriolis field.

The time-independent inhomogeneous Klein-Gordon equations
for the mesonic fields are given by
\begin{eqnarray}
\left\{-\Delta-(\Omega_x\hat{L}_x)^2+m_\sigma^2\right\}  
\sigma(\bff r)=-g_\sigma\left[\rho_s^p(\bff r)+\rho_s^n(\bff
r)\right]\qquad  &&
\nonumber \\  
-g_2\sigma^2(\bff r)-g_3\sigma^3(\bff r), \qquad  &&
\nonumber \\
\left\{-\Delta-(\Omega_x\hat{L}_x)^2+m_\omega^2\right\} 
\omega_0(\bff r) =
g_\omega\left[\rho_v^p(\bff r)+\rho_v^n(\bff r)\right], & &
\nonumber \\
\left\{-\Delta-(\Omega_x(\hat{L}_x+\hat{S}_x))^2+
m_\omega^2\right\} 
\bff\omega(\bff r) =
g_\omega\left[\bff j^p(\bff r)+\bff j^n(\bff r)\right], &&
\nonumber \\ \hspace{-5.0cm}  
\label{KGCRMF}
\end{eqnarray}
with source terms involving the various nucleonic densities
and currents
\begin{eqnarray}
\rho_s^{n,p}(\bff r) & = & \sum_{i=1}^{N,Z}(\psi_i(\bff r))^{\dagger} \hat{\beta}
\psi_i(\bff r), \nonumber \\
\rho_v^{n,p} (\bff r) & = & \sum_{i=1}^{N,Z} (\psi_i (\bff r))^{\dagger} \psi_i
(\bff r), \nonumber \\
\bff j^{n,p}(\bff r) & = & \sum_{i=1}^{N,Z} (\psi_i(\bff r))^{\dagger}
\hat{\bff\alpha} \psi_i (\bff r),
\label{curr}
\end{eqnarray}
where the labels $n$ and $p$ are used for neutrons and 
protons, respectively.  In the equations above, the sums run 
over the occupied positive-energy shell model states only 
({\it no-sea approximation}) \cite{NL1}. For simplicity, the 
equations for the $\rho$ meson and the Coulomb fields are 
omitted in Eqs.\ (\ref{KGCRMF}) since they have the structure 
similar to the equations for the $\omega$ meson. Since the 
coupling constant of the
electromagnetic interaction is small compared with the coupling
constants of the meson fields, the Coriolis term for the Coulomb
potential $A_0(\bff r)$ and the spatial components of the vector
potential $\bff A(\bff r)$ are neglected in the calculations.
Note that in the CRMF theory the currents $\bff j^{n,p} (\bff r)$ 
are the products of the large ($G$) and  small ($F$) components of 
the Dirac spinors $\psi_i (\bff r)$.

 The Coriolis operator $\hat{J}_x$ and the magnetic potential 
$\bff V(\bff r)$ in the Dirac equation as well as the currents 
$\bff j^{n,p}(\bff r)$ in the Klein-Gordon equations do not
appear in the RMF equations for time-reversal systems \cite{SW.86} 
since they break this symmetry \cite{Gbook}. Similar to the 
non-relativistic case \cite{RS.80}, the presence of the Coriolis 
operator leads to the appearance of the time-odd mean fields. 
However, the CRMF calculations with only these time-odd fields 
accounted underestimate the experimental moments of inertia by 20-30\% 
\cite{KR.93,AKR.96}. A similar situation holds also in non-relativistic
theories, see Refs.\ \cite{DD.95,YM.99} and references therein.
The inclusion of the currents $\bff j^{n,p}(\bff r)$, which leads to 
the space-like components of the vector $\omega$ and $\rho$ mesons 
and thus to magnetic potential $\bff V(\bff r)$, considerably improves 
the description of experimental moments of inertia. The effect coming 
from the space-like components of the vector mesons is commonly 
referred as {\it nuclear magnetism} (further NM) \cite{KR.89} in the 
framework of RMF theory.

 Before we discuss the microscopic impact of NM 
on the moments of inertia, we remind that the kinematic ($J^{(1)}$) 
and dynamic ($J^{(2)}$) moments of inertia are defined as 
\begin{eqnarray}
J^{(1)}(\Omega_x)=\frac{J}{\Omega_x}, \, \qquad
J^{(2)}(\Omega_x)=\frac{dJ}{d\Omega_x},
\label{J1J2}
\end{eqnarray}
where the rotational frequency $\Omega_x$ along the $x$-axis in 
a one-dimensional cranking approximation is determined by the 
condition that the expectation value of the total angular 
momentum $J$ at spin $I$ has a definite value \cite{Ing.54}:
\begin{eqnarray}
J=\langle{\Phi_\Omega}\mid\hat{J}_x
\mid {\Phi_\Omega}\rangle~=~\sqrt{I(I+1)} =
\sum_{i=1}^A\left\langle i\left|\hat{j}_x\right| i\right\rangle
\label{Jexpct}
\end{eqnarray}
Note that $J$ is defined in the cranking model as a sum of the 
expectation values of the single-particle angular momentum 
operators $\hat{j}_x$ of the occupied states. This suggests that 
the modifications of 
the moments of inertia due to NM should be traced to the 
changes of the single-particle $\langle \hat{j}_x \rangle_i$ 
values.

\begin{figure}
\epsfxsize 16.0cm
\epsfbox{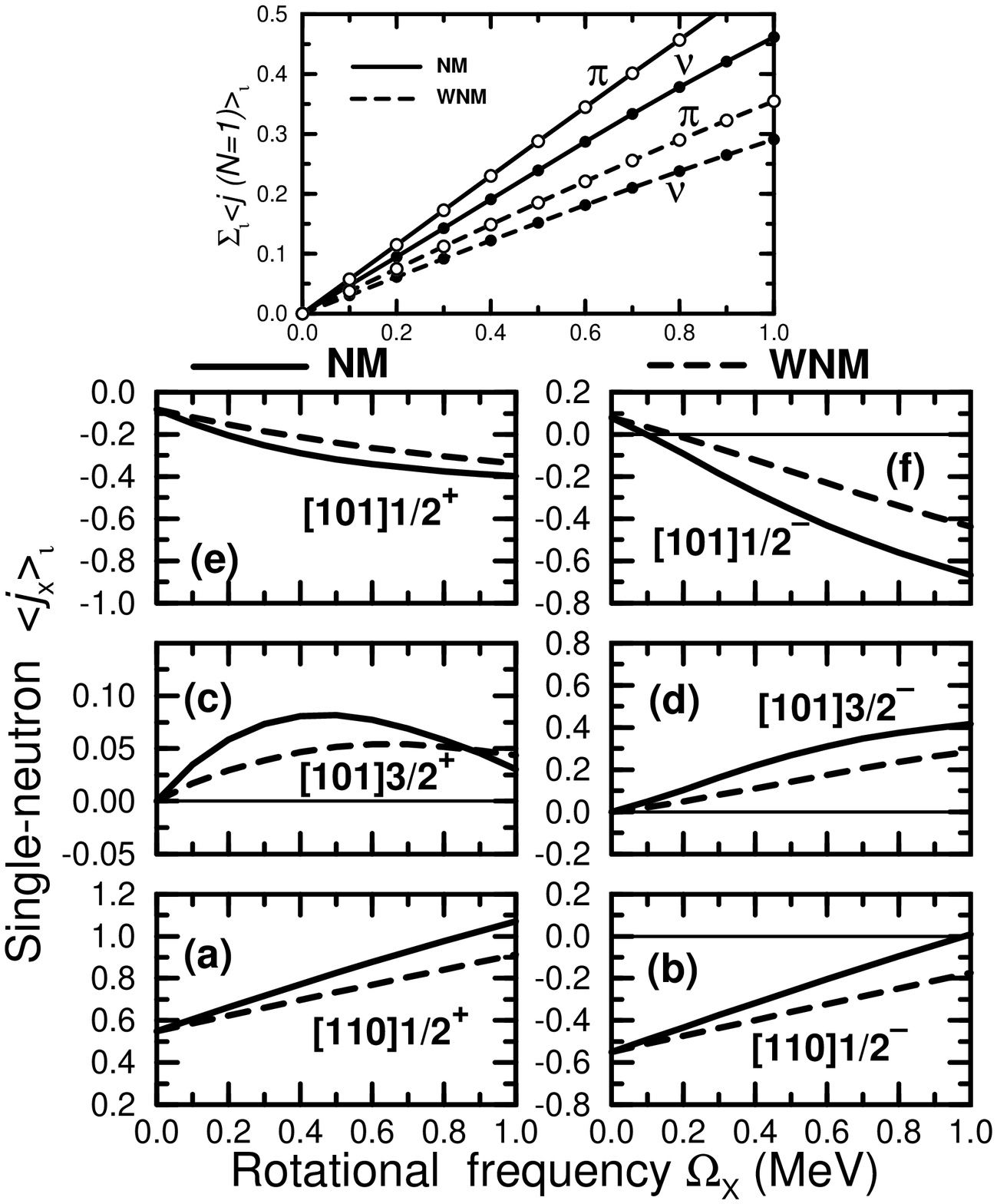}
\caption{The expectation values $ \langle j_x \rangle_i$ of the 
neutron orbitals forming the $N=1$ shell calculated with 
and without NM. They are given along the deformation
path of the lowest SD configuration in $^{152}$Dy.
While at $\Omega_x=0.0$ MeV the quadrupole moments
calculated with and without NM
differ by $\sim 10^{-4}$\%, this difference reaches 2.7\%
at $\Omega_x=1.00$ MeV. The notation of the lines is given in 
the figure. The single-particle orbitals are given from the 
bottom to the top according to their order within the $N=1$ shell, 
see Fig.\ \protect\ref{routh}. They are labeled by means
of the asymptotic quantum numbers $[Nn_z\Lambda]\Omega$ (Nilsson
quantum numbers) of their dominant component of the wave function
at $\Omega_x=0.0$ MeV and the sign of the signature quantum
number $r=\pm i$ given as superscript. Since the absolute values of 
$\langle j_x \rangle_i$ are small for the $[101]3/2^+$ orbital, the 
vertical scale of panel (c) is different from the other panels. 
In some panels, the horizontal thin lines with $\langle j_x \rangle
=0$ are shown in order to guide eye. Top panel shows the proton 
and neutron angular momenta $\sum_i <j_x(N=1)>_i$ generated by the 
orbitals of the $N=1$ shell. The results of the calculations with 
and without NM are displayed. Note that in the lowest SD configuration of 
$^{152}$Dy  the total $J$ values increase smoothly from 0 at 
$\Omega_x=0.0$ MeV to $88.26\hbar$ (NM) and $70.30\hbar$ (WNM) 
at $\Omega_x=1.0$ MeV.}
\label{N1shel}
\end{figure}

  In order to confirm this expectation the CRMF calculations with 
and without NM (the later will be further denoted as WNM) have been 
performed for the doubly magic superdeformed (SD) 
configuration in $^{152}$Dy ($\pi 6^4 \nu 7^2$) \cite{AKR.96}. 
This configuration has been selected due to the large stability 
of the SD minimum against rotation. The calculated 
$\langle \hat{j}_x \rangle_i$ values of the single-neutron orbitals 
forming the $N=1$ shell are given in Fig.\ \ref{N1shel}. This 
shell has been selected because (i) it is reasonably well separated 
in energy from the $N=3$ shell and thus the emerging picture is not 
strongly disturbed by an interaction between the $N=1$ and the 
$N=3$ shells, (ii) the number of single-particle orbitals is, from 
one hand, reasonably small to make an analysis transparent and, 
from other hand, sufficient to make meaningful conclusions.

Indeed NM has a considerable impact on the 
expectation values  of the single-particle angular momentum 
$\langle \hat{j}_x \rangle_i$. If we define the contribution 
to $\langle \hat{j}_x \rangle_i$ due to NM as 
\begin{eqnarray}
\Delta \langle j_x \rangle_i= \langle \hat{j}_x \rangle^{NM}_i 
- \langle \hat{j}_x \rangle^{WNM}_i
\label{Deltaj}
\end{eqnarray}
then one can see in Fig.\ \ref{N1shel} that $\Delta \langle j_x
\rangle_i$ is positive at the bottom and negative at the top 
of the $N=1$ shell. The absolute value of $\Delta \langle j_x)$
correlates with the absolute value of $\langle \hat{j}_x \rangle_i$. 
Note that the contributions to $\langle \hat{j}_x \rangle_i$ due to 
NM are small in the middle of the shell. It was checked also that 
similar features hold also in higher-$N$ shells. Note that the 
total angular momentum built within the $N=1$ shell 
is not equal 0 (see top panel in Fig.\ \ref{N1shel}) due to the 
admixture of the $N=3$ shell and that NM increases mixing of the 
$N=3$ and $N=1$ shells as seen from an increased value of this 
momentum and from the analysis of the structure of the wave 
functions of the single-particle orbitals. These results suggest 
that if it would be possible to isolate the $N$-shell then the 
contribution of NM to total angular momentum of this shell 
would be equal zero. 

\begin{figure}
\epsfxsize 16.0cm
\epsfbox{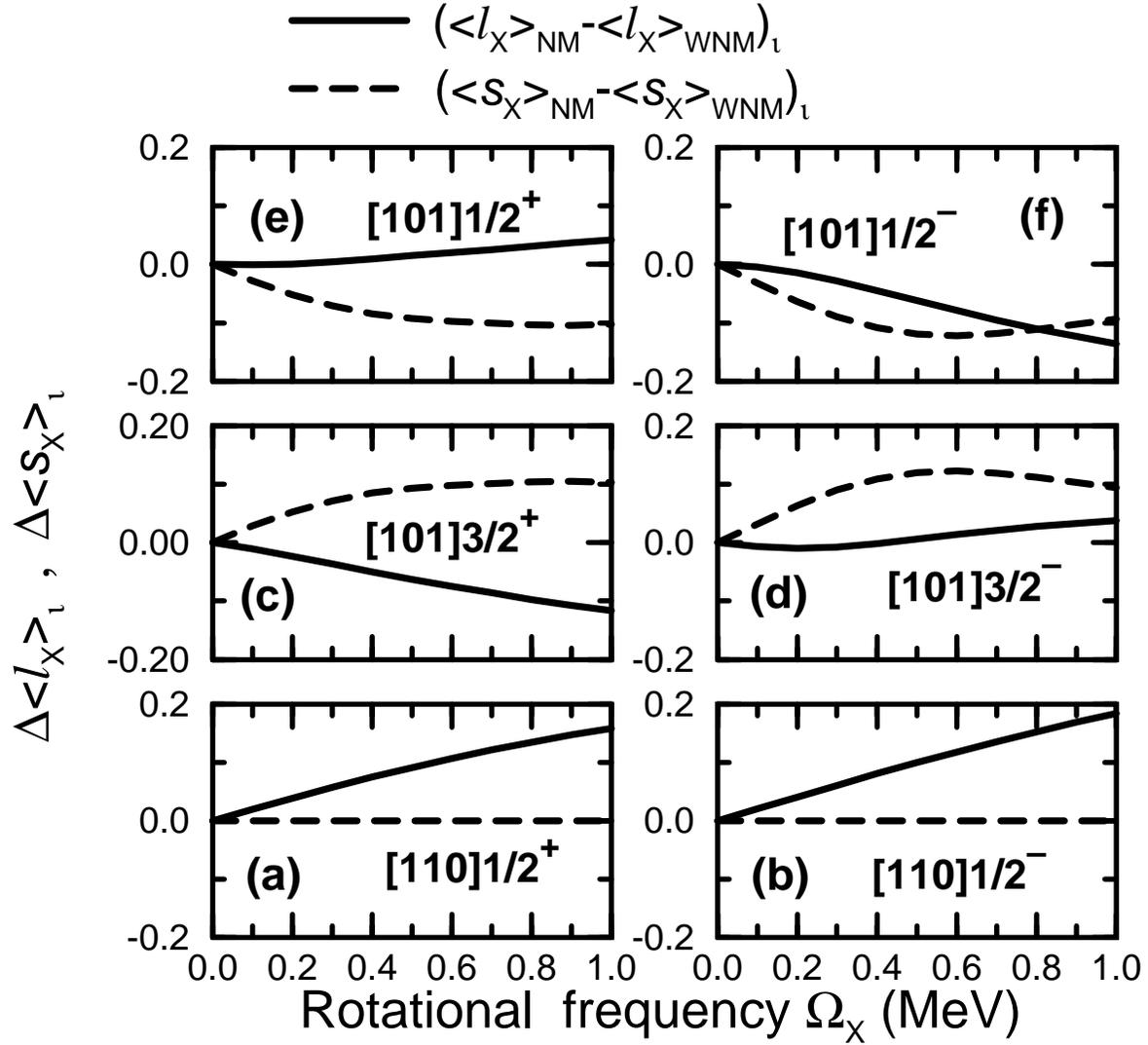}
\caption{The contributions to the expectation 
values of the spin and orbital angular momenta of 
the particles in the single-particle orbitals 
forming the $N=1$ shell caused by NM.}
\label{N1lxsx}
\end{figure}

 In order to gain a deeper understanding of the impact of NM, 
the angular momentum of the particle along the 
rotation axis is separated into the orbital and the spin parts: 
$\hat{j}_x=\hat{l}_x+\hat{s}_x$. Then the contributions due to 
NM to the expectation values of the orbital 
$(\Delta \langle \hat{l}_x \rangle_i)$ and the spin 
$(\Delta \langle \hat{s}_x \rangle_i)$ angular 
momenta, which are defined similar to Eq.\ (\ref{Deltaj}),
are plotted in Fig.\ \ref{N1lxsx}.

  The spin angular momenta of the particles in the $[110]1/2^+$ 
and $[110]1/2^-$ orbitals, which are the lowest orbitals in the
$N=1$ shell,  are almost fully aligned along the axis of rotation 
already at $\Omega_x=0.0$ MeV. Their expectation values 
$\langle \hat{s}_x \rangle_i$ are equal to $\sim 0.49\hbar$ and 
$\sim -0.49\hbar$ at all rotational frequencies employed. Thus 
NM cannot have an impact on the spin angular 
momenta of the particles in these orbitals 
($\Delta \langle \hat{s}_x \rangle_i\approx 0 \hbar$, 
see Figs.\ \ref{N1lxsx}a,b) and the contributions to 
$\langle \hat{j}_x \rangle_i$ due to NM
are almost solely defined by the modifications of the
expectation values of the orbital angular momentum 
$\langle \hat{l}_x \rangle_i$ induced by NM, see 
Figs.\ \ref{N1lxsx}a,b. On the contrary, the spin
angular momenta of the particles in the other orbitals 
forming the $N=1$ shell are only partially aligned along 
the rotation axis. Thus NM modifies 
the expectation values of both the spin and orbital angular 
momenta of the particles in these orbitals, see 
Figs.\ \ref{N1lxsx}c,d,e,f. Note that the contributions 
$\Delta \langle \hat{l}_x \rangle_i$ and 
$\Delta \langle \hat{s}_x \rangle_i$ have in general different 
rotational frequency dependence and can have different signs. 
The later is clearly seen in the case of the $[101]3/2^+$ 
orbital, see Figs.\ \ref{N1lxsx}c. The results for proton 
orbitals are very similar to the ones presented here for 
neutrons. Thus one can conclude that, in general,  NM leads to 
the modification of alignment properties of single-particle 
orbitals on the level of spin, orbital and total angular 
momenta.

  It was checked that the impact of NM on $\langle \hat{j}_x \rangle_i$
has the same general features also in other shells. However, for
higher $N$-shells the crossings between the single-particle 
routhians make the emerging picture more complicate. The question 
is then how to understand the increase of kinematic and dynamic
moments of inertia induced by NM. In the case of yrast SD band in 
$^{152}$Dy, the shells up to $N=3$ can be considered as 'closed' 
since all the orbitals belonging to them are occupied. 
The contribution of the 'closed' shells to the expectation value 
$J$ of the total angular momentum accounts only for few \% of $J$
as illustrated on the example of the $N=1$ shell, see top panel 
in Fig.\ \ref{N1shel}.  Thus the main contribution to $J$ (and
correspondingly to $J^{(1)}$ and $J^{(2)}$) is coming from the 
particles and holes outside the 'closed' shells. 

In the lowest SD configuration of $^{152}$Dy, valence 
protons are sitting in the orbitals, emerging from the bottom of 
the $N=5$ and $N=6$ shells (see for example Fig.\ 2 in Ref.\ 
\cite{TR.85}), for which the contribution of NM to $\langle 
\hat{j}_x \rangle_i$ is positive. Valence proton holes are 
located in the orbitals, emerging from the middle and the top 
of the $N=4$ shell. The contribution of NM to 
$\langle \hat{j}_x \rangle_i$ for these orbitals is either 
close to zero or negative. However, these contributions 
should be taken with opposite sign due to hole nature.
Thus these proton holes also give a positive contribution due 
to NM to the total proton angular momentum. The consideration 
of the neutron subsystem can be performed in a similar way and 
the conclusions are the same. Thus one can conclude that
at given rotational frequency the modifications of single-particle 
$\langle \hat{j}_x \rangle_i$  values induced by NM will lead to 
an increase of total $J$ and, as a result, to an increase of 
kinematic and dynamic moments of inertia. Essentially this 
means that NM is the source of the creation of angular 
momentum in rotating nuclei in addition to the Coriolis force.
Clearly this increase in $J$, $J^{(1)}$ and $J^{(2)}$ induced 
by NM should depend both on the rotational frequency and on the 
shell filling (and thus on the specific single-particle
configuration).

\begin{figure}
\epsfxsize 16.0cm
\epsfbox{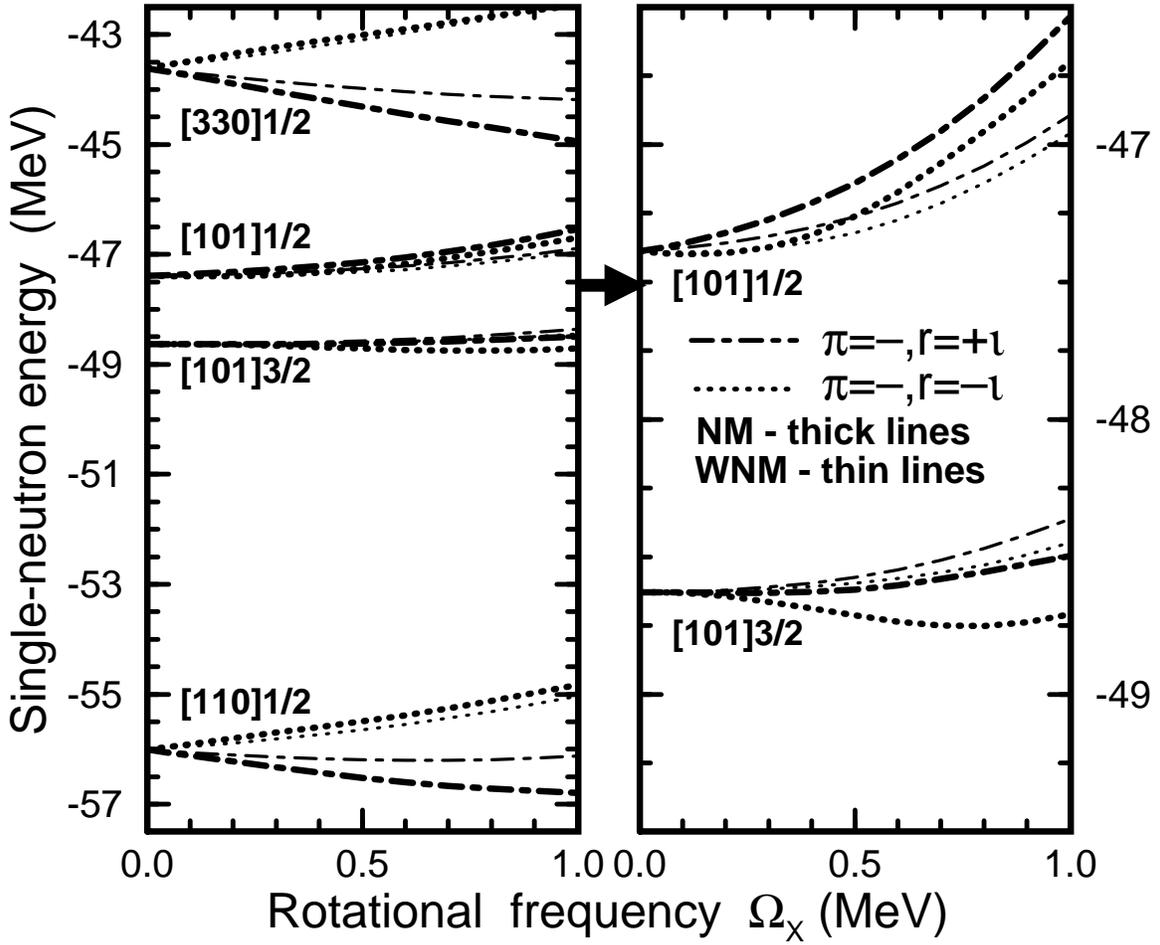}
\caption{Neutron single-particle energies (routhians) in the
self-consistent rotating potential, calculated with and
without nuclear magnetism, as a function of the rotational 
frequency $\Omega_x$. They are given along the
deformation path of the lowest SD configuration in
$^{152}$Dy. Only all the $N=1$ and the lowest $N=3$ orbitals 
are shown. The right panel shows the $[101]3/2$ and $[101]1/2$
orbitals in more details. The notation of the lines is given 
in figure.}
\label{routh}
\end{figure}

In the light of these findings, it is important to see which 
other physical quantities are affected by NM. NM can have considerable 
impact on the signature splitting of the signature partner orbitals 
as it is clearly seen in the case of the $[110]1/2^{\pm}$ and 
$[330]1/2^{\pm}$ orbitals, see Fig.\ \ref{routh}. One should also 
note that the change in $\langle \hat{j}_x \rangle_i$ induced by 
NM is not fully reflected in the change of the slope of the 
single-particle energies drawn as a function of the rotational 
frequency (see Fig.\ \ref{routh}). This is contrary to the results 
of the phenomenological models based on the Woods-Saxon or Nilsson 
potentials where for a cranking at a fixed deformation, the slope of 
the orbital routhian against rotational frequency corresponds to 
the  alignment 
$\langle \hat{j}_x \rangle_i=-\delta \epsilon_i^{\Omega_x}/\delta \Omega_x$,
where $\epsilon_i^{\Omega_x}$ is the single-particle energy of 
$i$-th orbital at given $\Omega_x$. For example, the 
$[110]1/2^{\pm}$ and $[101]1/2^-$ orbitals have the 
contributions to $\langle \hat{j} \rangle_i$ induced by 
NM (see Figs.\ \ref{N1shel}a,b,f) similar in absolute
value. However, the change of the slope of the single-particle 
routhians induced by NM is larger for the $[110]1/2^+$ orbital 
as compared with the $[110]1/2^+$ and $[101]1/2^-$ orbitals, 
see left panel of Fig.\ \ref{routh}. This essentially suggests 
that the difference between the single-particle energies
of the orbitals calculated with and without NM reflects
not only the change of alignment properties of corresponding
single-particle orbitals but also the energy gain (or cost) 
due to the interaction of the nucleon with the magnetic 
potential $\bff V (\bff r)$.

 In order to check that the features mentioned above are not coming 
from small differences in equilibrium deformations obtained in the 
calculations with and without NM, the calculations without NM have 
been performed starting from the self-consistent fields obtained 
in the calculations with NM and performing only one iteration
during which the physical quantities of interest have been 
recalculated neglecting NM. Although the solution obtained in this case
is not self-consistent, its equilibrium deformation differs 
only marginally from the one obtained in fully self-consistent 
calculations with NM. Using this result a similar analysis as 
presented above has been repeated. Although some minor quantitative 
differences exist, the physical conclusions are the same. 
 
\begin{figure}
\epsfysize 14.0cm
\epsfbox{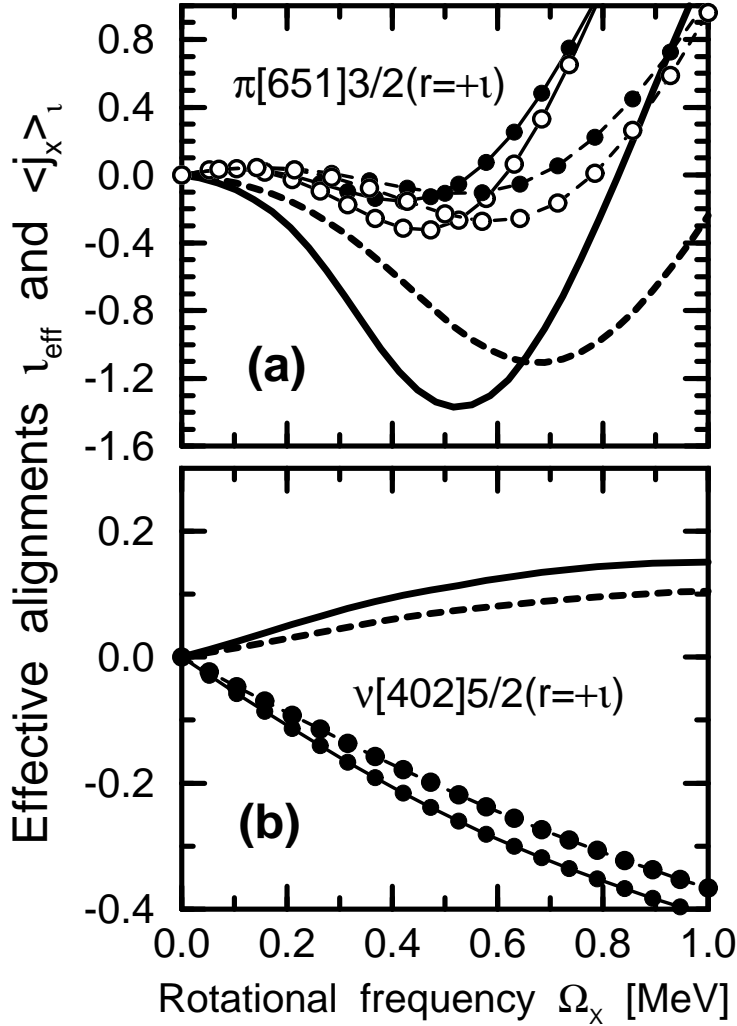}
\caption{The effective alignments (in $\hbar$) between 
the SD configurations in $^{151}$Tb and $^{152}$Dy
(panel (a)) and in $^{152}$Dy and $^{153}$Dy
(panel (b)) are shown by the lines without
symbols. The configurations in $^{151}$Tb and 
in $^{153}$Dy are obtained with respect to the lowest 
SD configuration in $^{152}$Dy by emptying of the 
$\pi [651]3/2(r=+i)$ orbital and by the occupation of 
the $\pi [402]5/2(r=+i)$ orbital, see Ref.\ 
\protect\cite{ALR.98} for details. The expectation 
values of the single-particle angular momenta 
$\langle j_x \rangle_i$ of these orbitals are 
shown by the lines with symbols. Solid and open
symbols are used for values calculated in 
$^{152}$Dy and the neighboring odd nucleus, 
respectively. Solid and dashed lines are used 
for the results obtained with and without NM. In 
panel (b), the $\langle \hat{j}_x \rangle_i$ values 
calculated in $^{153}$Dy are not shown since they differ 
from the ones obtained in $^{152}$Dy by less than 
$0.026\hbar$.}
\label{efalign}
\end{figure}

It is reasonable to expect that also the physical observables which 
depend on the alignment properties of specific single-particle
orbitals are affected by NM. One 
of these observables is the effective alignment $i_{eff}$ which
measures the difference between the spins of bands A and B
at constant $\Omega_x$ : 
$i_{eff}^{\rm A,B}(\Omega_x)=I_{\rm B}(\Omega_x)-I_{\rm A}(\Omega_x)$
and which in Cranked-Nilsson+Strutinsky (CNS) approach based on the 
cranked Nilsson potential is predominantly defined by the alignment(s) 
of the single-particle orbital(s) by which the compared bands differ 
\cite{R.93}. Fig.\ \ref{efalign} shows that NM indeed can have a 
considerable 
impact (especially in the case of the $\pi [651]3/2(r=+i)$ orbital) on 
effective alignments. A similar conclusion about the impact of time-odd
components of the mean field on relative (effective) alignments has been 
drawn earlier in the cranked Hartree-Fock approach based on Skyrme forces 
\cite{DD.95}. Note that contrary to the CNS approach polarization
effects play a much more important role in the CRMF theory and thus 
effective alignments are not defined predominantly by single-particle 
alignments, see Fig.\ \ref{efalign}. The fact that effective
alignments between SD bands in different mass regions  
\cite{ALR.98,A60,Hung} are correctly described 
in the CRMF theory indicates that both alignment properties of 
single-particle orbitals and relevant polarization effects are
properly accounted in this theory.

   In conclusion, the microscopic role of nuclear magnetism 
in rotating frame has been investigated for the first time. 
The breaking of time-reversal symmetry induced by rotation 
results in baryonic currents which lead to the non-vanishing 
space-like components of the vector mesons creating the magnetic 
potential entering into the Dirac equation. The magnetic field 
created by this potential modifies the expectation values of the 
single-particle angular momenta effectively creating additional
total angular momentum. Both the spin and the orbital single-particle 
angular momenta are affected by nuclear magnetism. This effect 
is state- and rotational frequency dependent. It leads to the 
increase of total angular momentum, kinematic and dynamic moments 
of inertia at given rotational frequency and has an impact on
signature splitting, single-particle energies and on effective 
alignments. It is reasonable to expect that an increase in 
kinematic and dynamic moments of inertia resulting from the 
time-odd components of the mean field in the non-relativistic 
mean field theories \cite{DD.95,YM.99} is caused by a similar 
microscopic mechanism as the one discussed in the present manuscript. 
Clearly more detailed studies of the microscopic nature of nuclear 
magnetism both in non-rotating and rotating cases are mandatory. 
Such an investigation is in progress and its results will be
reported later.

    A.V.A. acknowledges support from the Alexander von Humboldt 
Foundation. This work is also supported in part by the 
Bundesministerium f{\"u}r Bildung und Forschung under the 
project 06 TM 979.


\vspace*{-0.2cm}

\end{document}